\documentclass{elsart}
\input{epsf}
\usepackage{psfig}

\usepackage{natbib}

\begin{document}


\begin{frontmatter}

\title{The UV spectra of NLS1s -- Implications for their broad line regions}

\author[CfA]{J.K. Kuraszkiewicz}  
\author[CfA]{B.J. Wilkes}
\author[CAMK]{B. Czerny} 
\author[Ohio]{S. Mathur}  
\author[PSU]{W.N. Brandt}  
\author[Ohio]{M. Vestergaard}

\address[CfA]{Harvard-Smithsonian Center for Astrophysics, Cambridge,
MA 02138, USA}
\address[CAMK]{N. Copernicus Astronomical Center, Warsaw, Poland}
\address[Ohio]{The Ohio State University, Columbus, OH 43120, USA}
\address[PSU]{The Pennsylvania State University, University Park, 
PA 16802, USA}

\begin{abstract}

We study the UV spectra of NLS1 galaxies and compare them with typical
Seyfert~1 galaxies and quasars. The NLS1 spectra show narrower UV
lines as well as weaker CIV\,$\lambda$1549 and CIII]\,$\lambda$1909
emission. We show that these line properties are due to a lower
ionization parameter and somewhat higher BLR cloud densities. These
modified conditions can be explained by the hotter big blue bumps
observed in NLS1s, which are in turn due to higher $L/L_{Edd}$ ratios,
as shown by our accretion disk and corona modeling of the NLS1
continua.
We also present evidence that the Boroson \& Green eigenvector~1,
which is correlated with the optical and UV emission-line properties,
is not driven by orientation and hence NLS1s, which have extreme
eigenvector~1 values, are not viewed from an extreme viewing angle.

\end{abstract}


\end{frontmatter}


\section{UV emission lines}

\vspace{-0.5cm}

The optical spectra of NLS1s are characterized by narrow Balmer lines,
strong Fe~II\,$\lambda$4570 and weak [O~III]\,$\lambda$5007 (hereafter
[O~III]) emission. Here we study the UV spectra of NLS1s. Our sample
consists of 9 NLS1s for which UV spectra were available (as of
Dec. 1998) either from the HST or IUE archives. We found that the NLS1s
have narrower UV lines than typical broad-line AGN such as Seyfert~1
(Sy1) galaxies and quasars. The comparison of the equivalent widths of
the prominent UV lines in NLS1s showed that the distribution of
Ly$\alpha$ equivalent widths does not differ from typical AGN,
while the equivalent widths of CIV are significantly lower.

\begin{figure}[]
\psfig{figure=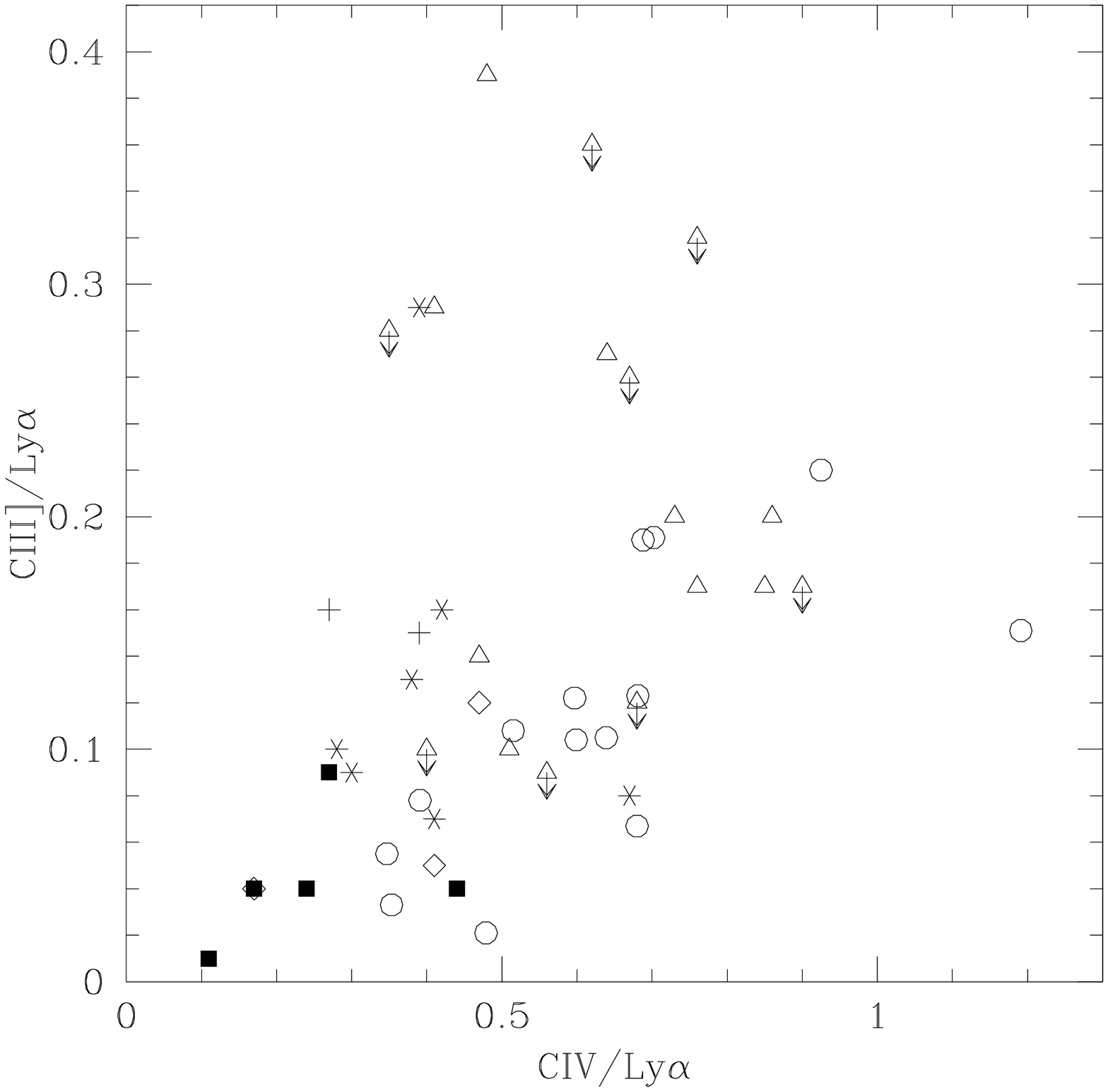,height=2.5in}
\vskip -2.5in
\hskip 2.9in
\psfig{figure=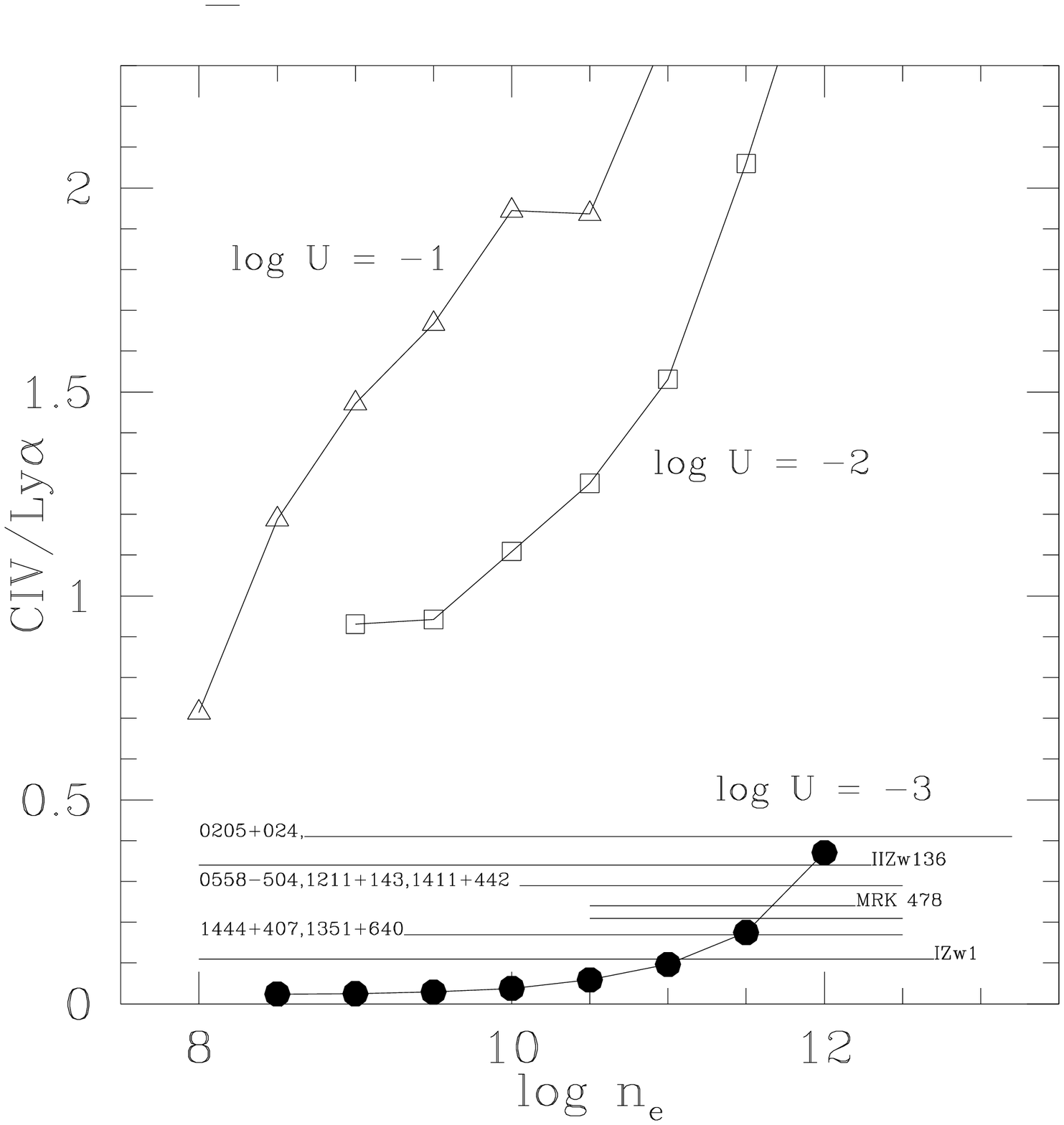,height=2.5in}
\caption{(Left): Comparison of CIII]/Ly$\alpha$ and CIV/Ly$\alpha$
ratios. NLS1s occupy the lower left corner of this plot (filled
squares). The rest of the data points are Seyfert~1s and quasars.}
\caption{(Right): Calculated CIV/Ly$\alpha$ ratio as a function of 
density ($n_e$) for different ionization parameters $U$. Horizontal 
lines show values of observed NLS1 line ratios. Low $U$ and high 
densities are clearly favored by the data. The observed range of
CIV/Ly$\alpha$ in typical AGN is 0.4--1}  
\end{figure}

The CIII]\,$\lambda$1909/Ly$\alpha$ and CIV/Ly$\alpha$ line ratios are
smaller (see Fig.~1) and the SiIII]\,$\lambda$1892/CIII] and
SiIV+OIV]\,$\lambda$1400/CIV ratios are larger when compared to
typical AGN.
To find what broad-line emitting cloud parameters produce such line
ratios, we have calculated these ratios using the ionization code {\it
CLOUDY} (Ferland 1991) for different densities and ionization
parameters, using as an input ionizing continuum the spectral energy
distribution of one of our NLS1s. An example of such a calculation for
the CIV/Ly$\alpha$ ratio is shown in Fig.~2.  In order to produce the
line ratios observed in NLS1s, 10 times lower an ionization parameter
($\log U = -3$) and densities up to 10 times larger (10$^{11}$~cm$^{-3}$ 
to 10$^{12}$~cm$^{-3}$ for CIV, Ly$\alpha$, SiIV emitting
clouds and $n_e = 10^{9.5}$~cm$^{-3}$ to 10$^{10.5}$~cm$^{-3}$ for the
SiIII], CIII] clouds) than standard are needed (where the standard BLR
parameters are $\log U = -2$ and $n_e = 10^{11}$~cm$^{-3}$ for the CIV,
Ly$\alpha$ emitting region and $n_e = 10^{9.5}$~cm$^{-3}$ for the
SiIII], CIII] region).

\section{NLS1 continuum properties}

\vspace{-0.5cm}

It is known (e.g. Puchnarewicz et al. 1995; Boller, Brandt \& Fink
1996) that the big blue bumps (BBBs) in NLS1s are shifted towards
higher energies compared to other AGN.  It has been suggested (Pounds,
Done \& Osborne 1995; Wandel 1997), based on the analogy with the
Galactic black hole candidates, that these hotter BBBs may be due to
higher ratios of the luminosity to the Eddington luminosity
($L/L_{Edd}$), meaning that NLS1s have systematically lower masses at
a given luminosity range than other AGN. We investigate this
suggestion by calculating the central engine parameters with the
accretion disk and corona model of Witt, Czerny \& \.Zycki (1997) and
fitting it to the optical/UV/X-ray continua of NLS1s. 
%
The model is fully defined by: the mass of the central black hole
($M_{bh}$), $L/L_{Edd}$ and the viscosity parameter ($\alpha_{vis}$).
We found that:
\vspace{-0.3cm}
 \begin{itemize}
\item $L/L_{Edd}$ is higher ($> 0.27$) in NLS1s than in 
typical AGN ($<$ 0.3)
\item NLS1s have 10--100 $\times$ lower $M_{bh}$ than AGN
with comparable luminosity 
\item bolometric luminosity of NLS1s is comparable to QSOs rather
than Sy1s.
\end{itemize}

The higher $L/L_{Edd}$ ratios result in hotter BBBs observed in NLS1s.
In the context of the two-phase model of Krolik \& Kallman (1988),
where the cool phase represents the BLR clouds and the hot phase
represents the intercloud confining medium, the hotter BBBs change the
equilibrium between the two phases. This results in higher BLR cloud
densities and hence the characteristic, observed UV emission-line
properties of NLS1s.


\section{Eigenvector~1 and orientation in radio-quiet quasars}

\vspace{-0.5cm}

It has also been suggested that another parameter, orientation, may be
responsible for the observed emission-line properties. Boroson \&
Green (1992, hereafter BG92) performed a principal component analysis
on the line/continuum correlation matrix of the Bright Quasar Survey
(BQS) sample and showed that the primary eigenvector (hereafter EV1),
was anticorrelated with various measures of Fe~II\,$\lambda 4570$
strength, correlated with [O~III] strength (luminosity and peak) and
H$\beta$ FWHM, and anticorrelated with the blue asymmetry of the
H$\beta$ line.  It was later found that these optical line properties
correlate with UV line properties and X-ray continuum
properties. NLS1s, as objects with strong Fe~II and weak [O~III] emission,
have the most negative values of EV1.

BG92 and Boroson (1992) argued that EV1 is not driven by an
orientation effect (i.e. some anisotropic property) as it is strongly
correlated with the [O~III] luminosity, which was assumed to be
isotropic.  However, the isotropy of the [O~III] emission in other AGN
has since been called into question. Some results based on radio-loud
samples (Jackson \& Browne 1990; Baker 1997) suggest that the inner
regions of [O~III] emission extend to sufficiently small radii to be
obscured by the dusty torus when the active nucleus is viewed
edge-on. On the other hand, [O~II]\,$\lambda$3727 emission, which has a
lower critical density and ionization potential, lies further out and
should not be affected by the torus (Hes, Barthel \& Fosbury
1996). Hence we present an investigation of the relation between 
[O~II] emission and EV1, which will enable us to find whether EV1 is
dependent on orientation. Finding a strong relation between EV1 and
[O~II] luminosity would imply that EV1 is independent of orientation,
while the lack of such a relation would suggest that orientation is a
factor.
 
We have chosen radio-quiet quasars from the optically selected Bright
Quasar Survey (BQS) from the BG92 sample covering a wide range of EV1
values. We have subtracted from our spectra the Fe~II emission, which
contaminates the [O~III] emission, and we have accounted for the presence
of the small blue bump (Balmer continuum and Fe~II emission) at the
[O~II] wavelengths.

We found strong correlations between L([O~II]), L([O~III]) and
EV1 (see Fig.~3), implying that EV1 does not depend
on orientation, confirming earlier conclusions of BG92 and Boroson
(1992), based on [O~III] alone.  It also suggests that both [O~II] and
[O~III] emission does not depend on orientation in BQS radio-quiet
quasars. This also implies that NLS1s, which lie at extreme
EV1 values, are not viewed from an extreme viewing angle.

\begin{figure}
\begin{center}
\leavevmode
\hbox{%
\epsfxsize=4.5in
\epsfbox{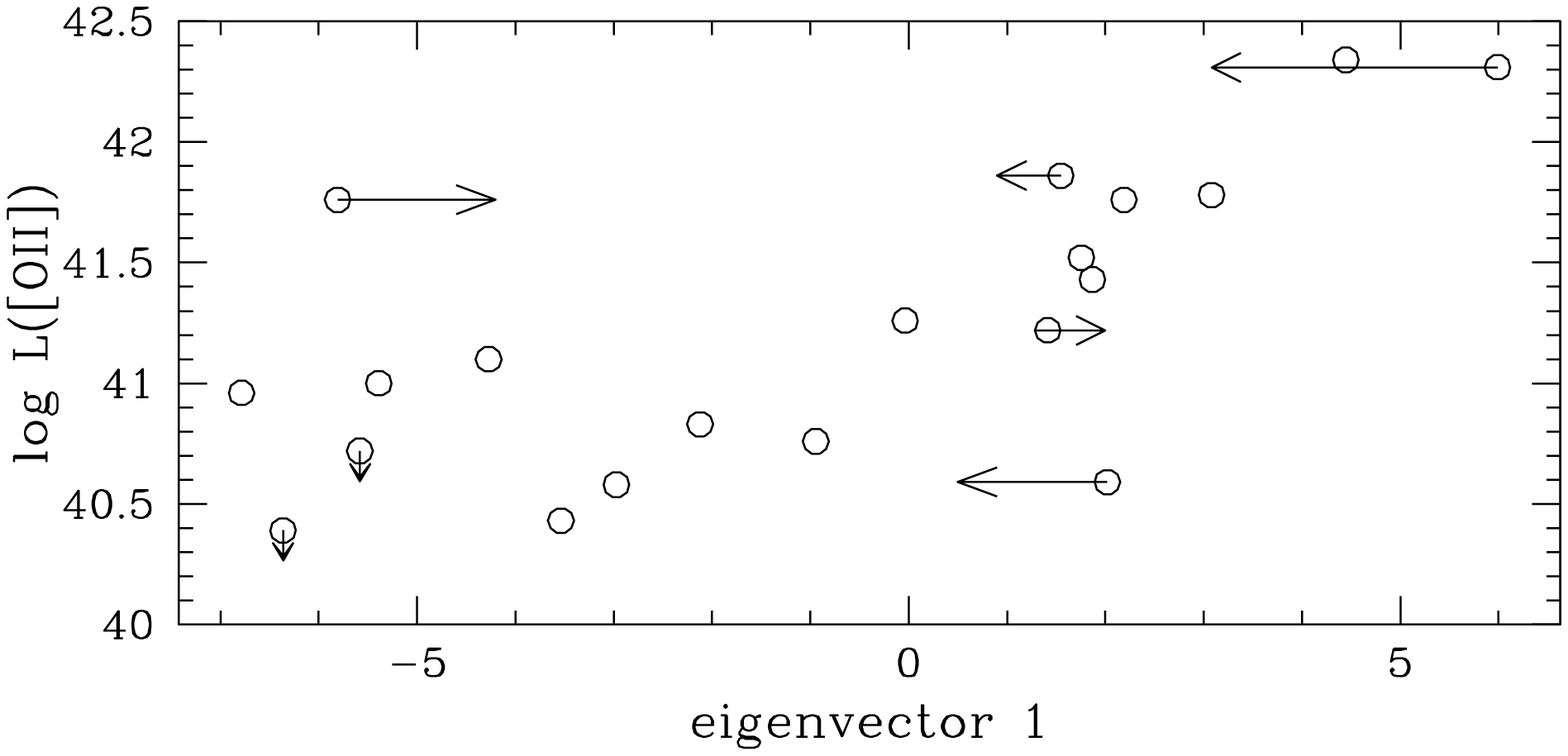}}
\end{center}
\vspace{-2.2in}
\caption{The [O~II] luminosity versus Boroson \& Green
eigenvector~1 correlation.}
\label{fig-3}
\end{figure}



\begin{thebibliography}{99}
\bibitem{1} Baker, J. C. 1997, MNRAS, 286, 2
\bibitem{2} Boller, Th., Brandt, W. N., \& Fink, H. 1996, AA, 305, 53
\bibitem{3} Boroson, T. A., \& Green, R. F. 1992, ApJS, 80, 109
\bibitem{4} Boroson, T. A.  1992, ApJ, 399, 15
\bibitem{5} Ferland, G. F. 1991, ``HAZY'', OSU Astronomy Dept.
Internal Rept. 
\bibitem{6} Hes, R., Barthel, P. D., \& Fosbury, R. E. A. 1996, A\&A, 313,
423
\bibitem{7} Jackson, N., \& Browne, I. W. A. 1990, Nature, 343, 43
\bibitem{8} Krolik, J. H., \& Kallman, T. R. 1988, ApJ, 324, 714
\bibitem{9} Pounds, K. A., Done, C., \& Osborne, J. P. 1995, MNRAS, 277, L5
\bibitem{10} Puchnarewicz, E. M., Mason, K. O., Siemiginowska, A., \&
Pounds, K. A. 1995, MNRAS, 276, 20
\bibitem{11} Wandel, A. 1997, ApJ, 490, L131
\bibitem{12} Witt, H. J. Czerny, B., \& Zycki, P. T. 1997, MNRAS, 286, 848
\end{thebibliography}
\end{document}